\documentclass[twocolumn,showkeys,preprintnumbers,amsmath,amssymb]{revtex4}
\usepackage{graphicx}
\usepackage{dcolumn}
\begin{document}

\title{Effects of decoherence on the radiative and squeezing properties in a coherently driven trapped
two-level atom}
\author{Sintayehu Tesfa}
\affiliation{Physics Department, Addis Ababa University, P. O. Box 1176, Addis Ababa, Ethiopia}%
\email{sint_tesfa@yahoo.com}

\date{\today}

\begin{abstract}Analysis of the effects of decoherence on the radiative and
squeezing properties of a coherently driven two-level atom trapped
in a resonant cavity applying the corresponding master equation is
presented. The atomic dynamics as well as the squeezing and
statistical properties of the emitted radiation are investigated.
It is found that the atom stays in the lower energy level more
often at steady state irrespective of the strength of the coherent
radiation and thermal fluctuations entering the cavity. Moreover, a
strong external coherent radiation results the splitting of the
line of the emission spectrum, whereas the decoherence broadens
the width and significantly decreases the height. It is also found
that the emitted radiation exhibits photon anti-bunching,
super-Poissonian photon statistics and squeezing, despite the
presence of the decoherence which is expected to destroy the
quantum features.
\end{abstract}

\keywords{two-level atom, atomic dynamics, quadrature squeezing,
emission spectrum,  decoherence}
 \maketitle

\section{Introduction}

Interaction of a single two-level atom with a radiation
has received a great deal of interest in recent years
\cite{jmo,jmo451859,prl561917,jmo34821,el10237,jobqso4142,jmo48347,jmo46379,pr1881969,pra414083,pra426873,pra373867,prl582539,oc118143,jmo431555,pra534439,pra533633,pra532846,pra74063817}.
In the absence of the external light, spontaneous emission of an
excited two-level atom results due to the fluctuations in a
continuum of vacuum modes that play a role of a reservoir.
However, when the vacuum modes are replaced with, let us say, a
squeezed vacuum reservoir, the radiative properties of the atom
are significantly modified
\cite{jmo,jmo451859,prl561917,jmo34821,el10237,jobqso4142,jmo48347,jmo46379}.
In this respect, the resonance fluorescence and absorption spectra
of a driven two-level atom coupled to a squeezed vacuum reservoir
have been analyzed by various authors \cite{jmo,jmo451859}. It is,
in general, found that the rate of emission is greater than the
rate of absorption, in which the squeezed input inhibits
absorption somehow and broadens the spectrum with decreasing
height. Moreover, whenever the atom is driven on resonance by a strong
external monochromatic laser beam, the structure of the atomic
energy level changes dramatically, that is, the atomic dynamics and
properties of the emitted radiation would be appreciably altered.
It is known for long that the resonance fluorescence spectrum
splits into Mollow triplets in the strong driving regime
\cite{pr1881969}. It is also a well-established fact that  the radiative
properties of the atom and squeezing properties of the emitted
radiation considerably depend on the amplitude of the driving
radiation.

It can be learned from earlier works that the
successively emitted photons from the two-level atom in the cavity
are correlated due to the atomic coherence induced by the driving
mechanism, whereby the emitted radiation is found to exhibit
nonclassical features. For instance, D'Souza {\it{et al.}}
\cite{pra414083} analyzed the quantum nature of the light emitted
by the two-level atom coupled to a squeezed vacuum reservoir in
the strong driving limit with the aid of atomic-dressed state
earlier. Based on the phase sensitivity of the Mandel's response
function, they claimed that the emitted light exhibits squeezing.
Most recently, the squeezing properties of the radiation emitted by
a coherently driven two-level atom coupled to a broadband squeezed
vacuum reservoir is studied using the variance of the
atomic-dipole operator in the normal ordering. Successively
emitted radiation turns out to be in squeezed state even in the
absence of the squeezed input for certain values of the
amplitude of the external radiation. In addition, although it was predicted earlier that the emitted
radiation exhibits sub and super-Poissonian photon statistics,
from the curve of the Mandel's response function \cite{pra414083}, recent analysis
based on the two-time second-order correlation function shows that it exhibits super-Poissonian photon statistics for
larger delayed time \cite{jmo}. It is also found that the two-time
second-order correlation function oscillates with the delayed time,
where the frequency of its oscillation increases with the
amplitude of the driving radiation, whereas its height decays fast
with the squeeze parameter.

In the actual experimental setting, the two-level atom in the cavity is
unavoidably coupled to the fluctuations in the surrounding
environment via the walls of the cavity. In general, the phenomenon in which
the quantum system losses its nonclassical features due to its
interaction with the environment is defined as decoherence. It is
not difficult to realize, therefore, that decoherence is basically
related to unbiased noise fluctuations in the modes of the
environment that able to interact with the system. Though various ways of including the effects of
decoherence are possible \cite{pra445401}, its contribution can be readily modeled as
thermal fluctuations of the walls of the cavity that can be taken
usually as thermal reservoir. It is a well-known fact that a squeezed
vacuum reservoir introduces a biased noise fluctuations to the
system, as a result it induces additional coherence, whereas
the thermal reservoir, on the other hand, adds decoherence into the system. In view of the contribution of the squeezed vacuum reservoir towards
the nonclassical features of the emitted radiation that have been
reported, it appears natural to ask how the radiative properties
of the atom as well as the squeezing and statistical properties of
the emitted radiation could possibly be modified by decoherence due to the
presumed thermal heating entering the cavity via the vibration of
the walls of the container? On the basis of the properties of unbiased noise fluctuations associated with the thermal heating, it seams reasonable to expect that the quantum features of the radiation would be degraded by the decoherence. The main task of this work is,
thereupon, devoted to investigate this basic issue. Earlier, effects of the thermal light as incoherent relaxation on the
collapse and revival as well as the photon anti-bunching have been
considered by Puri and Agarwal \cite{pra353433}. They
found that the oscillations of the collapse and revival become
more irregular with the intensity of the thermal radiation and the
thermal light characteristically destroys the photon anti-bunching
phenomenon.

In this communication, the effects of decoherence on a radiative,
squeezing and statistical properties of a coherently driven
two-level atom would be analyzed. It is a common knowledge that the effects of squeezed input are related to the amplitude and phase fluctuations of the reservoir modes. Nonetheless, which of these two would be predominant one of the issues in this work. To achieve this goal, the squeezed input is replaced with an unbiased thermal fluctuations whose mean phase fluctuations  are readily averaged out to be zero. In accordance to this, throughout, the results previously obtained for squeezed input elsewhere \cite{jmo} are compared with the effects of the thermal fluctuations so that which of the fluctuations in the reservoir modes would  actually be essential in bringing about a significant modification in a radiative and squeezing properties ie evident. Though methods from the
stochastic simulation of the Bloch equations in secular
approximation \cite{pra426873,pra373867} to diagonalizing the
coefficient matrix \cite{jobqso4142,jmo48347,pra414083} have been
used in previous contributions, the differential equations
associated with the expectation values of the atomic and energy
operators following from the master equation would be
simultaneously solved in view of the procedure recently applied.
It is believed that this approach helps in overcoming the
inevitable limitations corresponding to the approximations and
computer simulation frequently employed. Usually the effects of
the external coherent radiation either in a weak or strong driving
limit have been studied, but in here an arbitrary amplitude of the
driving radiation is taken. In particular,  the population
inversion, probability for the atom to be in the upper energy
level, emission spectrum, two-time second-order correlation
function and quadrature variance for the cavity radiation in terms
of the atomic polarization would be calculated.

\section{Atomic Dynamics}

It is a common knowledge that driving a two-level atom on
resonance by a coherent light amounts to pumping the two-level
atom continuously by an external laser beam whose frequency
matches with the atomic transition frequency. Treating the driving
radiation classically, the Hamiltonian describing the interaction
of two-level atom with a radiation in the rotating-wave and
electric-dipole approximations in the interaction picture can be
expressed as
\begin{align}\label{tla01}\hat{H}=i{\Omega\over2}\big(\hat{\sigma}_{+}-\hat{\sigma}_{-}\big),\end{align}
where $\Omega$ is the positive-real constant proportional to the
amplitude of the external coherent radiation,
$\hat{\sigma}_{+}\;(\hat{\sigma}_{-})$ is the creation
(annihilation) atomic operator defined as
$\hat{\sigma}_{+}=|a\rangle\langle b|$ and
$\hat{\sigma}_{-}=|b\rangle\langle a|$ in which $|a\rangle$ and
$|b\rangle$ represent the upper and lower atomic energy levels. It
is a well known fact that the master equation of a two-level atom
coupled to a thermal reservoir can be derived applying the
Born-Markov approximation. Hence following the
 standard procedure \cite{lou}, it is possible to verify for a two-level atom driven on resonance
 by a
 coherent light and coupled to a
thermal reservoir that
\begin{align}\label{tla02}\frac{d\hat{\rho}}{dt}& =
\frac{\Omega}{2}\big(\hat{\sigma}_{+}\hat{\rho} -
\hat{\rho}\hat{\sigma}_{+}-\hat{\sigma}_{-}\hat{\rho}
 -\hat{\rho}\hat{\sigma}_{-}\big) \notag\\&+
\frac{\gamma\big(\bar{n}+1\big)}{2}\big[2\hat{\sigma}_{-}\hat{\rho}\hat{\sigma}_{+}
- \hat{\sigma}_{+}\hat{\sigma}_{-}\hat{\rho} -
\hat{\rho}\hat{\sigma}_{+}\hat{\sigma}_{-}\big] \notag\\&+
\frac{\gamma\bar{n}}{2}\big[2\hat{\sigma}_{+}\hat{\rho}\hat{\sigma}_{-}
- \hat{\sigma}_{-}\hat{\sigma}_{+}\hat{\rho} -
\hat{\rho}\hat{\sigma}_{-}\hat{\sigma}_{+}\big],\end{align} where
$\gamma$ is the atomic damping constant and $\bar{n}$ is the mean
photon number corresponding to the reservoir modes, which is the
measure of the intensity of the unbiased noise fluctuations of the
broadband environment modes.

Making use of the master equation \eqref{tla02}, the time
evolution of the expectation values of the atomic creation,
annihilation and energy operators can be obtained,
\begin{align}\label{tla03}\frac{d}{dt}\langle\hat{\sigma}_{-}(t)\rangle & =
 -
\frac{\gamma}{2}\big(2\bar{n} + 1\big)
\langle\hat{\sigma}_{-}(t)\rangle-\frac{\Omega}{2}\langle\hat{\sigma}_{z}(t)\rangle,
\end{align}
\begin{align}\label{tla04}\frac{d}{dt}\langle\hat{\sigma}_{+}(t)\rangle & =
 -
\frac{\gamma}{2}\big(2\bar{n} + 1\big)
\langle\hat{\sigma}_{+}(t)\rangle-\frac{\Omega}{2}\langle\hat{\sigma}_{z}(t)\rangle,
\end{align}
\begin{align}\label{tla05}\frac{d}{dt}\langle\hat{\sigma}_{z}(t)\rangle & =
 - \gamma\big(2\bar{n} +
1\big)\langle\hat{\sigma}_{z}(t)\rangle
\notag\\&+\Omega\big(\langle\hat{\sigma}_{-}(t)\rangle +
\langle\hat{\sigma}_{+}(t)\rangle\big)- \gamma.\end{align}
Following the procedure outlined in Ref. \cite{jmo}, it is
possible to show that
\begin{align}\label{tla06}\langle\hat{\sigma}_{z}(t)\rangle & =
\left(\langle\hat{\sigma}_{z}(0)\rangle +
\frac{\gamma^{2}(1+2\bar{n})}{2\alpha\beta}\right)e^{-\beta t} -
\frac{\gamma^{2}(1+2\bar{n})}{2\alpha\beta} \notag\\&+
\left[\frac{\beta - \gamma\big(2\bar{n} + 1\big)}{\beta -
\alpha}\langle\hat{\sigma}_{z}(0)\rangle  +
\frac{\gamma^{2}(1+2\bar{n})}{2\alpha\big(\beta -
\alpha\big)}\right.\notag\\&\left.+ \frac{\Omega}{\beta -
\alpha}\big(\langle\hat{\sigma}_{-}(0)\rangle +
\langle\hat{\sigma}_{+}(0)\rangle\big)- \frac{\gamma}{\beta -
\alpha}\right]\notag\\&\times\big(e^{-\alpha t} - e^{-\beta
t}\big),\end{align} where
\begin{align}\label{tla07}\alpha = \frac{\gamma}{4}(6\bar{n} +
3) - \xi,\end{align}
\begin{align}\label{tla08}\beta &= \frac{\gamma}{4}(6\bar{n} + 3) +
\xi,\end{align}
 in which
\begin{align}\label{tla09}\xi & =
\left[\frac{\gamma^{2}}{16}\big(2\bar{n} + 1\big)^{2} - \Omega^{2}
\right]^{1/2}.\end{align}
 It may worth mentioning that in the
forthcoming discussions various quantities of interest can be
determined using Eq. \eqref{tla06}.

Applying Eqs. \eqref{tla06}, \eqref{tla07}, \eqref{tla08},
\eqref{tla09}
 and the fact that the population inversion,
$W(t)=\langle\hat{\sigma}_{z}(t)\rangle$,  it is found  at steady
state that
\begin{align}\label{tla10}W=-{1
\over(1+2\bar{n})\left({2\Omega^{2}\over\gamma^{2}}+1\right)}.\end{align}
\begin{center}
\begin{figure}[hbt]
\centerline{\includegraphics [height=6cm,angle=0]{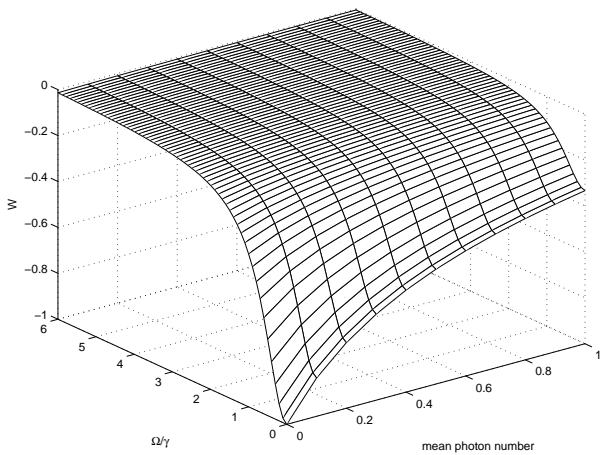}}
\caption{Plot of the atomic inversion at steady state. }
\end{figure}
\end{center}
The population inversion is defined as the difference of the
population in the lower and upper energy levels,
$\rho_{aa}-\rho_{bb}$. From the result shown in Fig. 1, it is not
difficult to observe that the population inversion increases with
the amplitude of the external coherent radiation and intensity of the thermal
fluctuations. Moreover, further scrutiny reveals that the
inversion decreases with the intensity of the thermal fluctuations
for larger values of the amplitude of the coherent radiation. As clearly shown in our previous work \cite{jmo}, the
population inversion decreases with the increasing degree of
squeeze parameter for larger values of $\Omega/\gamma$. However, comparison of the dependence of the population inversion on the intensity of the noise fluctuations in the two systems indicates that the squeezed input slightly enhances the decrement of the population inversion. Since the unbiased noise fluctuations in the thermal vibrations are presumed to be washed out in the process of calculating the mean values and hence $\bar{n}$ accounts for the intensity of the fluctuations alone. This is one of the essential differences in atomic
dynamics in cases of biased or unbiased noise fluctuations are
allowed to enter the cavity. It can also be inferred from this
result that the atom stays more often in the lower energy level at
steady state, since the population inversion is found to be
negative for all values of the parameters under consideration.

Furthermore, on the basis of the fact that the probability for
 the atom to be in the upper energy level is given by
 $\rho_{aa}(t)= {\langle\hat{\sigma}_{z}(t)\rangle+1\over2}$
 and
making use of Eq. \eqref{tla06} one gets at steady state
\begin{align}\label{tla11}\rho_{aa} =
\frac{{\Omega^{2}\over\gamma^{2}}+\bar{n}(1+2\bar{n})}{{2\Omega^{2}\over\gamma^{2}}+(2\bar{n}+1)^{2}}.\end{align}

\begin{center}
\begin{figure}[hbt]
\centerline{\includegraphics [height=6cm,angle=0]{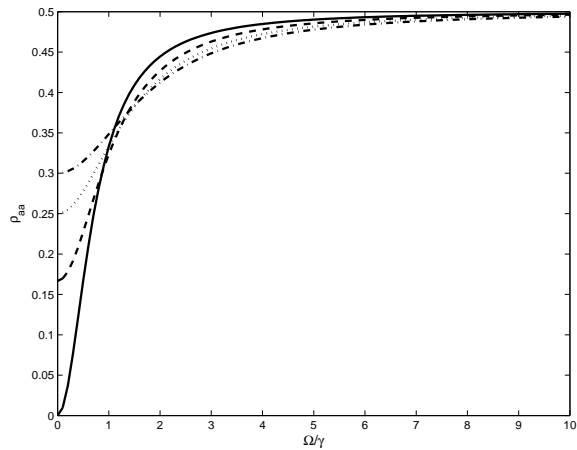}}
\caption {Plots of the probability for the atom to be in the upper
atomic energy level at steady state for $\bar{n}=0$ (solid line),
$\bar{n}=0.25$ (dashed line), $\bar{n}=0.5$ (dotted line) and
$\bar{n}=0.75$ (dashed-dotted line). }
\end{figure}
\end{center}
It can readily be seen from Fig. 2 that the probability for the
atom to be in the upper energy level increases with the mean
photon number of the reservoir modes for smaller values of
$\Omega/\gamma$, but decreases for larger values. In relation to
this, Tanas {\it{et al.}} \cite{jmo451859}  recently found that
the absorption spectrum of the driving field or the stationary
line shape, the quantity which is twice of the value in Eq.
\eqref{tla11}, is less than 1 at resonance when a strongly driven
two-level atom is coupled to a finite band squeezed vacuum
reservoir. The dependence of the
probability for the atom to be in the upper energy level on the
squeezed parameter has also the same form as indicated in Fig. 2. It,
hence, can be inferred that when there is an external radiation
the atom may absorb a photon from the cavity and then excited to
the upper energy level. Even then irrespective of the strength
of the external radiations (coherent driving and thermal
fluctuations) the rate of emission is relatively stronger than the
rate of absorption. As can readily be seen, $\rho_{aa}$ takes
values between 0 and 0.5, which implies that this mechanism
perhaps be employed in preparing the atom in an arbitrary
coherent superposition of the upper and lower energy levels by
adjusting the required amplitude of the external coherent
radiation. Despite previous claim that inhibition of absorption is
related to the phase difference between the coherent and squeezed
radiations, the result shown in this Section, rather, indicates
that the inhibition of absorption is predominantly depends on the
intensity of the fluctuations. It is good to note that there is a slight variation due to the phase sensitivity of the noise of
course.

\section{Emission Spectrum}

Emission spectrum that corresponds to the power spectrum of a
radiation emitted by  a two-level atom can be conveniently
expressed  in terms of the atomic creation and annihilation
operators as
\begin{align}\label{tla12}S(\omega)=2Re\int_{0}^{\infty}\langle\hat{\sigma}_{+}(t)\hat{\sigma}_{-}
(t+\tau)\rangle_{ss}e^{i\omega\tau}d\tau.\end{align}
Following the approach in Ref. \cite{jmo} along with the aid of
the properties of the atomic operators that
$\langle\hat{\sigma}^{2}_{-}\rangle=0$ and
$\langle\hat{\sigma}_{+}\hat{\sigma}_{z}\rangle=-\langle\hat{\sigma}_{+}\rangle$,
one can obtain
\begin{align}\label{tla13}&\langle\hat{\sigma}_{+}(t)\hat{\sigma}_{-}(t +
\tau)\rangle  =
\langle\sigma_{+}(t)\sigma_{-}(t)\rangle\left\{2e^{-\frac{\gamma}{2}(2\bar{n}+1)\tau}
\right.\notag\\&\left.-
\frac{\Omega^{2}}{2\left(\frac{\gamma(1+2\bar{n})}{2} -
\beta\right)(\beta-\alpha)}(e^{-\beta\tau} - e^{-\alpha\tau})\right\}
 \notag\\&+\langle\hat{\sigma}_{+}(t)
 \rangle\left\{\frac{-\Omega}{2\left(\frac{\gamma(1+2\bar{n})}{2} - \beta\right)}\left[
 \frac{\gamma}{\beta - \alpha} - \frac{\gamma^{2}(1+2\bar{n})}{2\beta(\beta - \alpha)} \right.\right.\notag\\&\left.\left.+ {\alpha -
 \gamma(2\bar{n}+1)\over\beta-\alpha}\right]e^{-\beta\tau} - \frac{\Omega}{2\left(\frac{\gamma(1+2\bar{n})}
 {2} - \alpha\right)}\left[\frac{\gamma^{2}(1+2\bar{n})}
 {2\alpha(\beta - \alpha)} \right.\right.\notag\\&\left.\left.-\frac{\gamma}{\beta -
\alpha} + \frac{\gamma(2\bar{n}+1)-\beta}{\beta -
\alpha}\right]e^{-\alpha\tau} \right.\notag\\&\left.+
\frac{\Omega\gamma^{2}}{2\beta\alpha}\left(1 -
e^{-\frac{\gamma(1+2\bar{n})}{2}\tau}\right)\right\},\end{align}
\begin{align}\label{tla14}\langle\hat{\sigma}_{+}\rangle_{ss}={\Omega\gamma^{2}\over2\alpha\beta}.\end{align}

 In order to
study the dependence of the emission spectrum on the amplitude of
the coherent radiation and intensity of the thermal fluctuations
more closely, two cases of interest are considered. For a strong
driving field, $\Omega \gg \gamma$, it is possible to easily see
from Eq. \eqref{tla09} that $\xi = i\Omega$, as a result, $\beta -
\alpha = i2\Omega$, $\frac{c\gamma}{2} - \beta = -i\Omega$,
$\frac{c\gamma}{2} - \alpha = i\Omega$, $\alpha\beta =
\Omega^{2}$, and $\langle\sigma_{+}(t)\rangle_{ss} = 0$. Moreover,
since
$\rho_{aa}(t)=\langle\hat{\sigma}_{+}(t)\hat{\sigma}_{-}(t)\rangle$
Eq. \eqref{tla11} reduces for $\Omega\gg\gamma$ and modest values
of $\bar{n}$ to
\begin{align}\label{tla15}\langle\sigma_{+}(t)\sigma_{-}(t)\rangle_{ss} =
\frac{1}{2}.\end{align}  It can be realized that, at steady state,
the population is independent of the strength of the decoherence
which is consistent with the result shown in Fig. 2. This would
strengthen the already established fact that to prepare a
two-level atom in a possible maximum coherent superposition of the
two energy levels, driving it with a strong external coherent
radiation is sufficient. Furthermore, it is not difficult to see
that
\begin{align}\label{tla16}\langle\hat{\sigma}_{+}(t)\hat{\sigma}_{-}(t +
\tau)\rangle_{ss} &=
\frac{1}{4}e^{-\frac{\gamma}{2}(2\bar{n}+1)\tau} \notag\\&+
\frac{1}{8}e^{(i\Omega -\frac{\gamma}{4}(6\bar{n} + 3))\tau}
\notag\\&+ \frac{1}{8}e^{-(\frac{\gamma}{4}(6\bar{n}+3) +
i\Omega)\tau},\end{align}
 from which follows
\begin{align}\label{tla17}S(\omega) &=
\frac{\frac{\gamma}{16}(6\bar{n}+3)}{(\Omega + \omega)^{2} +
[\frac{\gamma}{4}(6\bar{n}+3)]^{2}} \notag\\&+
\frac{\frac{\gamma}{16}(6\bar{n}+3)}{(\Omega - \omega)^{2} +
[\frac{\gamma}{4}(6\bar{n}+3)]^{2}} \notag\\&+
\frac{\frac{\gamma}{4}(1+2\bar{n})}{\omega^{2} +
[\frac{\gamma}{2}(1+2\bar{n})]^{2}}.\end{align}
\begin{center}
\begin{figure}[hbt]
\centerline{\includegraphics
[height=6cm,angle=0]{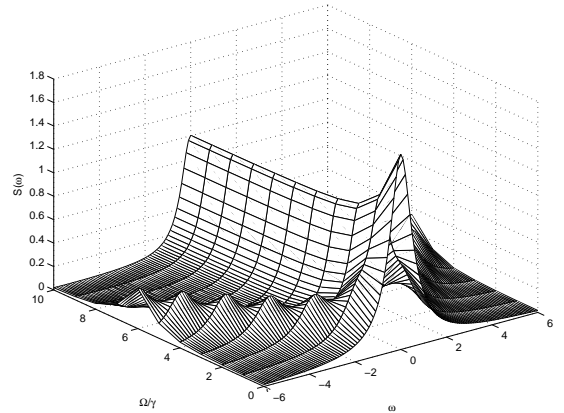}}\caption {Plot of the emission
spectrum at steady state for $\bar{n}=0.5$.}
\end{figure}
\end{center}

\begin{center}
\begin{figure}[hbt]
\centerline{\includegraphics
[height=6cm,angle=0]{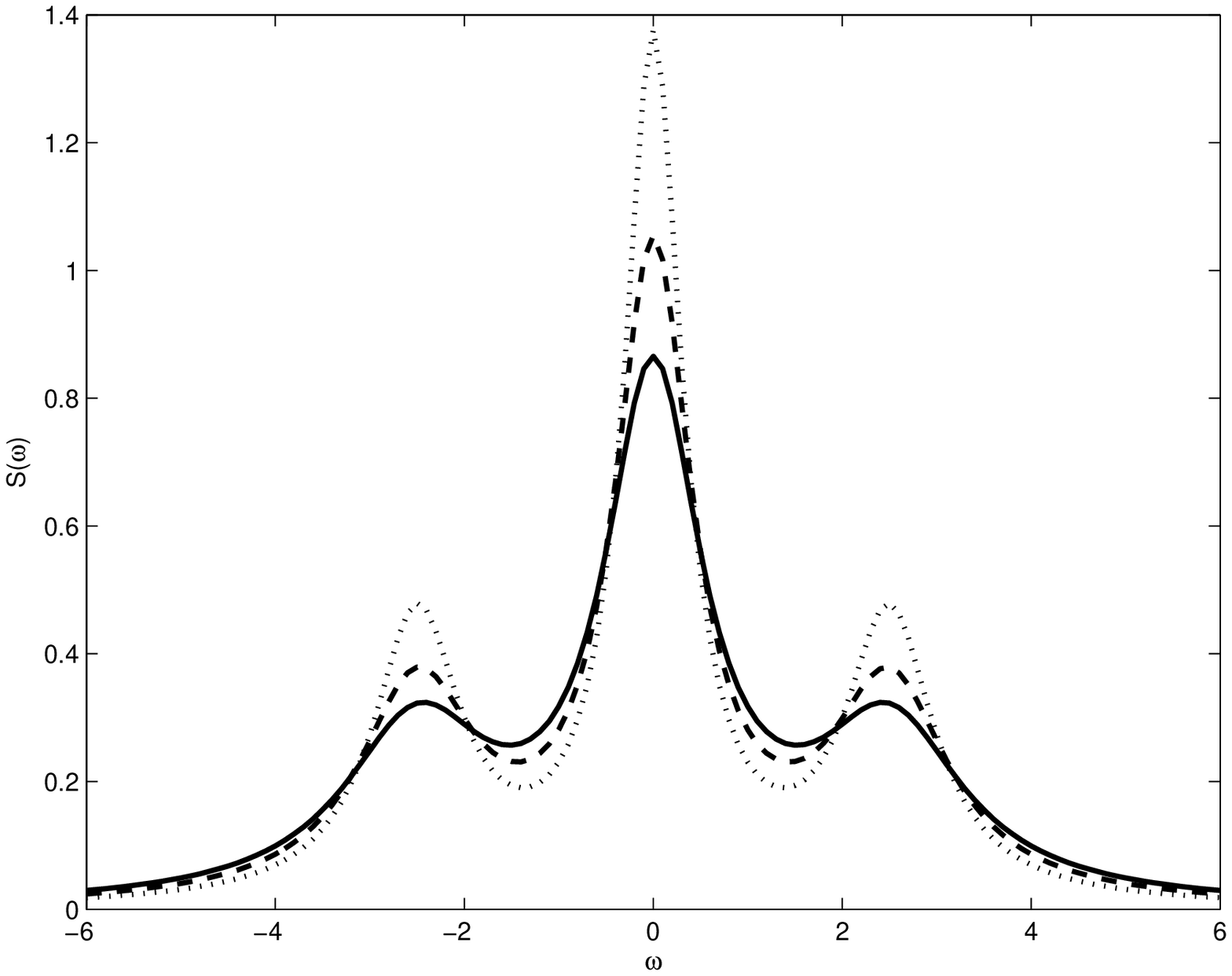}}\caption {Plots of the emission
spectrum at steady state for $\Omega=2.5\gamma$, $\bar{n}=0.75$
(solid line), $\bar{n}=0.5$ (dashed line) and $\bar{n}=0.25$
(dotted line).}
\end{figure}
\end{center}

\begin{center}
\begin{figure}[hbt]
\centerline{\includegraphics
[height=6cm,angle=0]{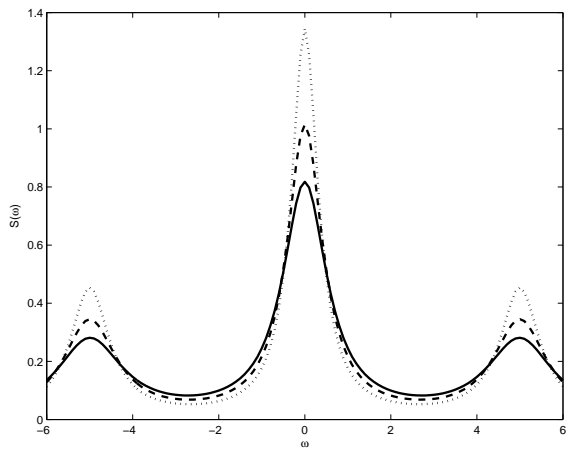}}\caption {Plots of the emission
spectrum at steady state for $\Omega=5\gamma$, $\bar{n}=0.75$
(solid line), $\bar{n}=0.5$ (dashed line) and $\bar{n}=0.25$
(dotted line).}
\end{figure}
\end{center}

It is not difficult to observe that the emission spectrum has
three well defined peaks at $\omega = 0$ and $\omega = \pm\Omega$
with line width of $\frac{\gamma(1+2\bar{n})}{2}$ and
$\frac{\gamma(6\bar{n} +3)}{4}$. For $\bar{n}=0$, this result goes
over to the usual Mollow type resonant fluorescent spectrum
\cite{pr1881969}. As can easily be seen from Fig. 3, the stronger
the intensity of the thermal fluctuations of the noise, the wider the splitting
and the shorter the height of the spectrum would be. In addition,
comparison of the results given in Figs. 4 and 5 shows that the
width of the central line and sidebands broadened with the
intensity of the decoherence, whereas the heights decreased. In
connection to this, Parkins \cite{pra426873} has simulated the
resonance fluorescence of a two-level atom coupled to a two-mode
squeezed vacuum reservoir and found that all the three peaks
exhibit subnatural line widths for particular choice of the phase
in a strong driving limit for a moderate squeezed input. On the
other hand, Tanas {\it{et al.}} \cite{jmo451859} recently
  found that the spectral lines of the resonance fluorescence of the
  two-level atom coupled to finite
band squeezed vacuum reservoir are narrower than for the ordinary
vacuum and the side bands are slightly shifted. It is now evident that the profile of the spectra are the same as what is obtained here even when the biased noise is replaced by unbiased noise fluctuations. Nonetheless, comparison with
previous results shows that the biased fluctuations in the
squeezed vacuum modes suppress the height of the central peak
prominently than the unbiased noise fluctuations in the
decoherence phenomenon, which is basically related to the phase
sensitivity in the squeezed input. On the basis of this understanding, one can then come to conclude that except for such minor
differences, the essential mechanism in emission-absorption process depends
on the intensity of the fluctuations of the noise associated with
the environment rather than the phase.

In  a weak driving limit, $\Omega\approx0$, it follows from Eq.
\eqref{tla13} that
\begin{align}\label{tla18}\langle\hat{\sigma}_{+}(t)\hat{\sigma}_{-}(t +
\tau)\rangle_{ss} & =
2\langle\hat{\sigma}_{+}(t)\hat{\sigma}_{-}(t)\rangle_{ss}
\left[e^{-\frac{\gamma}{2}(2\bar{n}+1)\tau}\right],\end{align}
which leads, making use of Eq. \eqref{tla11} at steady state,
$\langle\hat{\sigma}_{+}(t)\hat{\sigma}_{-}(t)\rangle_{ss}  =
\frac{\bar{n}}{2\bar{n}+1}$, to
\begin{align}\label{tla19}\langle\hat{\sigma}_{+}(t)\hat{\sigma}_{-}(t+\tau)\rangle_{ss}  =
\frac{\bar{2n}}{2\bar{n}+1}\left[e^{-\frac{\gamma}{2}(2\bar{n}+1)\tau}\right].\end{align}
It is, hence, observed that in the weak driving limit the noise
associated to the thermal fluctuations can excite the atom to the
upper energy level, namely, for strongly intense thermal light
there is nearly 50\% probability for the atom to be found in the
upper atomic energy level at steady state. Just like the coherent
driving radiation, the thermal fluctuations entering the cavity
through the walls of the mirror can also be employed in preparing
the atom in arbitrary coherent superposition of the two atomic
energy levels. One can easily see that the atom would be
completely in the ground state at steady state for vacuum
reservoir.
Moreover, it can be deduced from Eq. \eqref{tla19} that for a weak driving limit, $\Omega=0$, the emission spectrum
generally does not split. Therefore, it is possible to infer that the spectral splitting is associated with the strength
of the external coherent radiation, whereas broadening of the
width with the intensity of the fluctuations entering the cavity.

\section{Photon statistics of the cavity radiation}

Currently available literatures indicate that the photon
statistics of the cavity radiation can be investigated using the
normalized two-time second-order correlation function that can be
expressed for the two-level atom in terms of the creation and
annihilation atomic operators in the form
\begin{align}\label{tla21}g^{(2)}(\tau)={\langle\hat{\sigma}_{+}(t)
\hat{\sigma}_{+}(t+\tau)\hat{\sigma}_{-}(t+\tau)\hat{\sigma}_{-}(t)\rangle\over
\langle\hat{\sigma}_{+}(t)\hat{\sigma}_{-}(t)\rangle^{2}}.\end{align}
Therefore, in view of the property of the atomic operators,
\begin{align}\label{tla22}\langle\hat{\sigma}_{+}(t+\tau)\hat{\sigma}_{-}(t+\tau)\rangle={
\langle\hat{\sigma}_{z}(t+\tau)\rangle+1\over2},\end{align} one
gets
\begin{align}\label{tla23}g^{(2)}(\tau)={\langle\hat{\sigma}_{+}(t)\hat{\sigma}_{-}(t)\rangle
+\langle
\hat{\sigma}_{+}(t)\hat{\sigma}_{z}(t+\tau)\hat{\sigma}_{-}(t)\rangle\over2
\langle\hat{\sigma}_{+}(t)\hat{\sigma}_{-}(t)\rangle^{2}},\end{align}
from which follows
\begin{align}\label{tla24}g^{(2)}(\tau) & =
{1\over2\langle\hat{\sigma}_{+}(t)\hat{\sigma}_{-}(t)\rangle}
\left[{2\alpha\beta-\gamma^{2}(1+2\bar{n})\over2\alpha\beta}\right.\notag\\&\left.
+{2\beta\alpha-\gamma^{2}(1+2\bar{n})-4\beta\gamma
\bar{n}\over2\beta(\beta-\alpha)}e^{-\beta\tau}\right.\notag\\&\left.+{\gamma^{2}(1+2\bar{n})
-2\beta\alpha+4\alpha\gamma
\bar{n}\over2\alpha(\beta-\alpha)}e^{-\alpha\tau}\right].\end{align}
For a strong driving limit one finds at steady state
\begin{align}\label{tla25}g^{(2)}(\tau) & =
1- \cos(\Omega\tau)e^{-{\gamma\over4}(6\bar{n}+3)\tau}.\end{align}
\begin{center}
\begin{figure}[hbt]
\centerline{\includegraphics [height=6cm,angle=0]{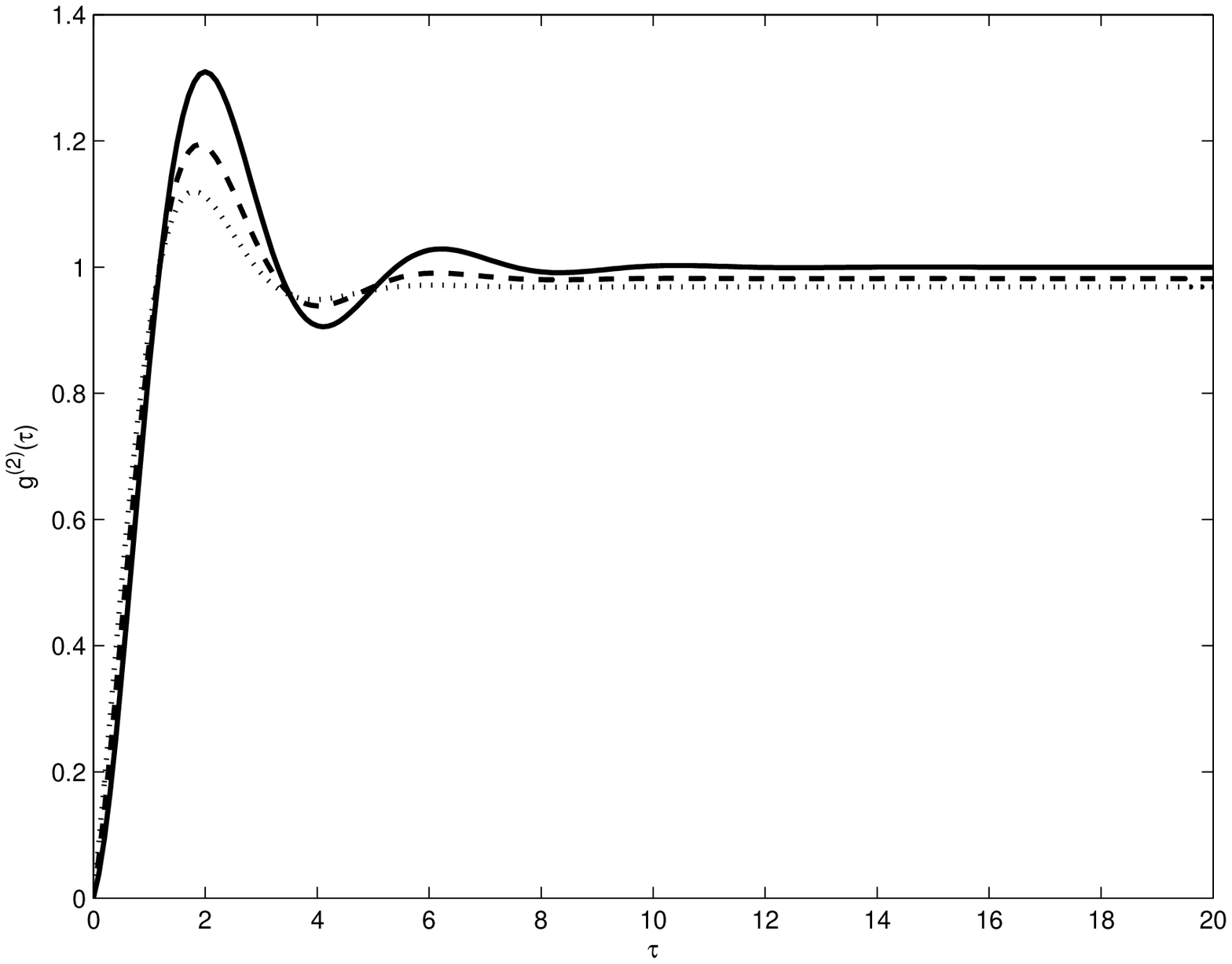}}
\caption {Plots of the two-time second-order correlation function
(Eq. \eqref{tla24}) at steady state for $\Omega=3\gamma$,
$\bar{n}=0.75$ (solid line), $\bar{n}=0.5$ (dashed line) and
$\bar{n}=0.25$ (dotted line).}
\end{figure}
\end{center}

\begin{center}
\begin{figure}[hbt]
\centerline{\includegraphics [height=6cm,angle=0]{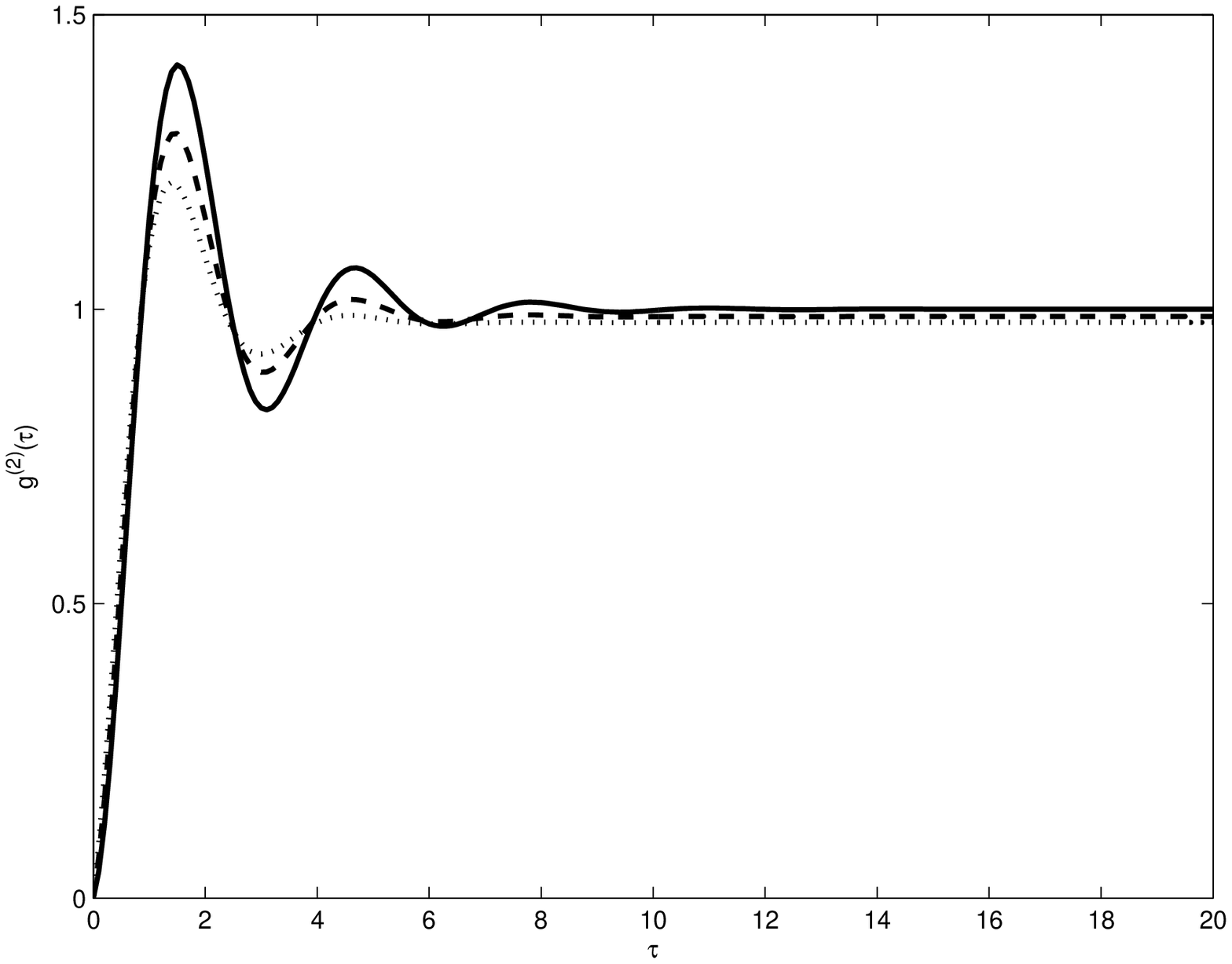}}
\caption {Plots of the two-time second-order correlation function
(Eq. \eqref{tla24}) at steady state for $\Omega=4\gamma$,
$\bar{n}=0.75$ (solid line), $\bar{n}=0.5$ (dashed line) and
$\bar{n}=0.25$ (dotted line).}
\end{figure}
\end{center}

\begin{center}
\begin{figure}[hbt]
\centerline{\includegraphics [height=6cm,angle=0]{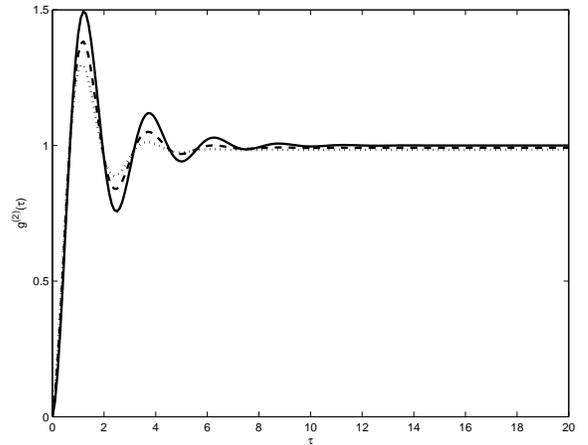}}
\caption {Plots of the two-time second-order correlation function
(Eq. \eqref{tla24}) at steady state for $\Omega=5\gamma$,
$\bar{n}=0.75$ (solid line), $\bar{n}=0.5$ (dashed line) and
$\bar{n}=0.25$ (dotted line).}
\end{figure}
\end{center}

It is known for long that the two-time second-order correlation
function describes the delayed coincidence between the
successively emitted light. It is not difficult to see from Figs.
6, 7 and 8 that $g^{(2)}(\tau)>g^{(2)}(0)$ for all cases under
consideration, which indicates that the emitted light exhibits
photon anti-bunching, despite the fact that the unbiased noise
fluctuations entering the cavity destroys the quantum features of
the radiation. The anti-bunching phenomenon can be interpreted as
the atom goes over to the lower energy level after emitting a
photon needs time before it absorbs a photon and excited to the
upper energy level to emit the next photon. It can be deduced that
this is one of the fundamental properties of the absorption and
emission processes of the two-level atom, which is independent of
the external coherent radiation and reservoir to which the cavity
is coupled. It is also possible to identify the photon statistics
of the cavity radiation employing the normalized two-time
second-order correlation function. As can readily be seen from
Figs. 6, 7 and 8, the two-time second-order correlation function
oscillates between $g^{(2)}(\tau)>1$ and $g^{(2)}(\tau)<1$ for
smaller values of the delayed time. This can be interpreted as the
photon statistics oscillates between sub and supper-Poissonian in
this case. However, for larger delayed time there is a possibility
that $g^{(2)}(\tau)>1$ or $g^{(2)}(\tau)<1$ depending on the
amplitude of the coherent radiation and the strength of the
decoherence. The super-Poissonian photon statistics becomes more
prominent for stronger intensity of decoherence. As can be
observed from Eq. \eqref{tla25} the emitted photon exhibits
Poissonian photon statistics for modest values of the amplitude of
the external radiation and for larger delayed time. Similar
oscillatory nature of the two-time second-order correlation
function with delayed time has been discussed by various authors
\cite{jmo,pra414083}. In particular, D'Souza {\it{et al.}}
\cite{pra414083} earlier predicted that the emitted radiation
exhibits both sub and super-Poissonian photon statistics based on
the phase between the coherent and squeezed lights from the curve
of the Mandel's response function, whereas recent study shows that
for larger delayed time the cavity radiation exhibits a
super-Poissonian photon statistics when the cavity is coupled to a
broadband squeezed vacuum reservoir \cite{jmo}. It is evident from
these works that the photon statistics at larger delayed time is
dominated by the properties of the light entering the cavity.

\section{Squeezing of the cavity radiation}

The squeezing properties of the cavity radiation can be studied
applying the variances of the atomic-dipole operators in the
normal order. In order to determine the variances of the
atomic-dipole operators in the normal order, it is possible to
begin with a well established fact that the emitted radiation can
be described in terms of the electric field. If the two quadrature
components of the electric field satisfy the commutation relation,
\begin{align}\label{tla26}\big[\hat{E}_{\theta},\;\hat{E}_{\theta-\pi/2}\big]=i2C,\end{align}
then the usual uncertainty relation,
\begin{align}\label{tla27}\langle(\Delta\hat{E}_{\theta})^{2}\rangle\langle(\Delta
\hat{E}_{\theta-\pi/2})^{2}\rangle\ge C^{2},\end{align} holds.
The radiation represented by this electric field is in squeezed
state, provided that either
$\langle(\Delta\hat{E}_{\theta})^{2}\rangle$ or $\langle(\Delta
\hat{E}_{\theta-\pi/2})^{2}\rangle$ is below the vacuum limit $C$.
In general, one of the variances of the electric field can be put
in the normal order as
\begin{align}\label{tla28}\langle:(\Delta\hat{E}_{\theta})^{2}:\rangle
=\langle(\Delta \hat{E}_{\theta})^{2}\rangle- C,\end{align} where
the symbol :: stands for the operator put in the normal order.
Therefore, the squeezing can be related to the requirement that
either $\langle:(\Delta\hat{E}_{\theta})^{2}:\rangle$ or
$\langle:(\Delta \hat{E}_{\theta-\pi/2})^{2}:\rangle$ is less than
zero. On the other hand, making use of the relation between the
electric field and atomic operators, the variance in the field
operator can be defined in terms of the atomic-dipole operators.
In this regard, Ficek and Tanas \cite{pr372369} have expressed the
variance of the atomic-dipole operator in the normal order in the
form
\begin{align}\label{tla29}\langle:(\Delta\hat{\sigma}_{i})^{2}:\rangle
=\langle(\Delta
\hat{\sigma}_{i})^{2}\rangle+{\langle\hat{\sigma}_{z}\rangle\over2},\end{align}
where $i=x,y$ and
\begin{align}\label{tla30}\hat{\sigma}_{x}={1\over\sqrt{2}}\big(\hat{\sigma}_{+}+\hat{\sigma}_{-}\big),\end{align}
\begin{align}\label{tla31}\hat{\sigma}_{y}={i\over\sqrt{2}}\big(\hat{\sigma}_{-}-\hat{\sigma}_{+}\big).\end{align}
Then with the aid  of the fact that at steady state
$\langle\hat{\sigma}_{-}(t)\rangle_{ss}=\langle\hat{\sigma}_{+}(t)\rangle_{ss}$,
one finds
\begin{align}\label{tla32}\langle:(\Delta\hat{\sigma}_{x})^{2}:\rangle
={1\over2}\big(1+\langle\hat{\sigma}_{z}\rangle\big)-2\langle\hat{\sigma}_{+}\rangle^{2}_{ss},\end{align}
\begin{align}\label{tla33}\langle:(\Delta\hat{\sigma}_{y})^{2}:\rangle
={1\over2}\big(1+\langle\hat{\sigma}_{z}\rangle\big).\end{align}

\begin{center}
\begin{figure}[hbt]
\centerline{\includegraphics [height=6cm,angle=0]{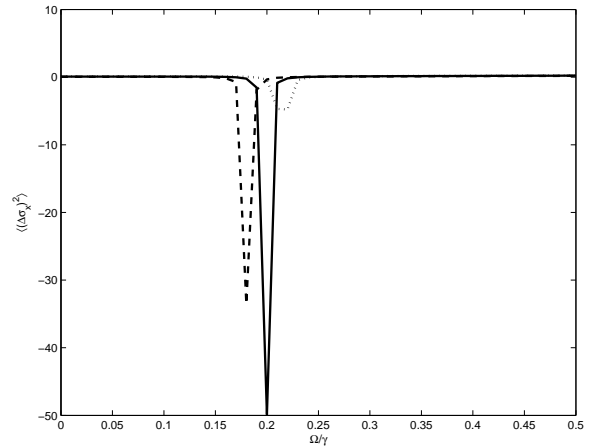}}
\caption {Plots of the squeezing of the cavity radiation at steady
state for $\bar{n}=0.05$ (dotted line), $\bar{n}=0.1$ (solid line)
and $\bar{n}=0.15$ (dashed line). }
\end{figure}
\end{center}

It is not difficult to realize based on the definition of the
population inversion along with the result shown in Fig. 1 that
$\langle\Delta\hat{\sigma}_{z}\rangle>-1$, which implies that
$\langle:(\Delta\hat{\sigma}_{y})^{2}:\rangle$ never be negative
at steady state. As can also be seen from Eq. \eqref{tla14},
$\langle\hat{\sigma}_{+}\rangle_{ss}=\gamma^{2}/2\Omega$, which
approaches zero for strong driving radiation. In this case as
well, it can readily be seen that
$\langle:(\Delta\hat{\sigma}_{x})^{2}:\rangle$ never be negative.
This indicates that the emitted radiation does not exhibit
squeezing when the atom is pumped externally with a strong
coherent radiation at steady state. As opposed to this,
$\langle:(\Delta\hat{\sigma}_{x})^{2}:\rangle$ can be less than
zero for certain values of the amplitude of the external radiation
as clearly shown in Fig. 9 for smaller values of $\Omega/\gamma$.
On the basis of the criterion set for squeezing, the light
emitted by the two-level atom exhibits squeezing even in the
presence of a significant amount of decoherence that is believed
to destroy the quantum features of the light. Unfortunately, the
squeezing is found to exist only for narrow pockets of the values
of the amplitude of the coherent radiation which, of course,
depend on the strength of the decoherence. This result
demonstrates that the atomic coherence induced between the upper
and lower energy levels by the coherent radiation, which is
responsible for the squeezing, is too strong to be destroyed by
the unbiased noise fluctuations. It is believed that this must be
the reason for observing a considerable entanglement in a
correlated emission laser even in the presence of a strong
decoherence \cite{jpbamop402373}. Moreover, critical survey of
Fig. 9 reveals that the squeezing exists for values of
$\Omega/\gamma$ for which the squeezing disappears in the absence
of the decoherence. This may be related to a recent claim that a
decoherence due to environment enhances entanglement in a
two-level atomic system by providing an indirect correlation
between totally uncorrelated quantum states \cite{pra75012101}.

\section{Conclusion}

In this contribution, a thorough study of the effects of the
external coherent radiation and thermal fluctuations corresponding
to the vibration of the walls of the mirrors due to their coupling
with the external environment on the atomic dynamics, squeezing
properties and photon statistics of the radiation produced by a
coherently driven  two-level atom trapped in a resonant cavity is
presented. It is found that though the atom absorbs the radiation
from the available cavity modes, including the driving, emitted
and thermal light entering the cavity, and makes a transition to
the upper energy level, it prefers to stay in the ground state
more often irrespective of the amplitude of the coherent radiation
and the strength of the intensity of the decoherence. In this
regard, in comparison to previous works, the fundamental
phenomenon in absorption and emission processes, in which the rate
of emission is greater than the rate of absorption, is basically found to
depend on the strength of the fluctuations associated to the
environment rather the phase difference. This, on the other hand,
indicates that except for the minor differences in its degree the
inhibition of absorption is resulted when the cavity is coupled to
both biased and unbiased noise fluctuations. Therefore, it is
possible to deduce from this study that predominantly the atomic
dynamics is affected by the mean photon number of the reservoir
modes rather than whether the reservoir is squeezed vacuum or
thermal. In addition to this, it is not difficult to realize that
the two-level atom can be prepared in arbitrary coherent
superposition of the upper and lower energy levels by varying the
intensity of the thermal fluctuations in the environment. It is
believed that this approach perhaps would be practically
attractive in the preparation of the injected atomic coherence required in
multi-level atomic laser \cite{jpbamop402373}.

It was previously discussed that the emission spectrum is
broadened and the height is reduced by the squeezed input. In the
same way, the thermal fluctuation is found to broaden the spectrum
and reduce the height significantly, but it does not contribute to
the splitting of the central line into triplet. Comparison with
the previous works indicates that the biased noise fluctuations in
the squeezed vacuum modes decrease the height of the central peak
more than the unbiased noise fluctuations in decoherence.
Moreover, the emitted radiation is found to exhibit anti-bunching
photon statistics independent of the type of the reservoir. It is,
rather, a fundamental property of a driven two-level atom related
to the time required for absorbing a radiation to make a
transition to the upper energy level after the atom emits a
photon. As opposed to this, the super-Poissonian statistics is
found to be enhanced by biased noise fluctuations. In addition to
this, the cavity radiation exhibits appreciable squeezing for some
pockets of the amplitude of the driving radiation that depends on
the strength of the intensity of decoherence. In conclusion, even
though the successively emitted photons are separated in time,
they are strongly correlated which leads to the appearance of the nonclassical
features even in the presence of decoherence which presumably
destroys the quantum properties.

\end{document}